\documentclass
[preprint,pre,10pt,twocolumn,superscriptaddress,notitlepage,showpacs,nofootinbib]{revtex4}%
\usepackage{amsfonts}
\usepackage{amsmath}
\usepackage{amssymb}
\usepackage{graphicx}%
\begin{document}
\title{Thermo-Statistical description of the Hamiltonian non extensive systems\\The selfsimilarity scaling laws}
\author{L. Velazquez}
\email{luisberis@geo.upr.edu.cu}
\affiliation{Departamento de F\'{\i}sica, Universidad de Pinar del Rio, Marti 270, Esq. 27
de Noviembre, Pinar del Rio, Cuba}
\author{F. Guzman}
\affiliation{Instituto Superior de Tecnologia y Ciencias Aplicadas, Quinta de los Molinos,
Plaza de la Revolucion, La Habana, Cuba}

\begin{abstract}
The foundations for a thermo-statistical description of the called non
extensive Hamiltonian systems are reconsidered. The relevance of the
parametric resonance as a fundamental mechanism of the Hamiltonian chaoticity
in those systems with bound motions in the configurational space is discussed.
The universality of this mechanism suggests us the possibility of performing a
thermo-statistical description with microcanonical basis in the context of the
long-range interacting Hamiltonian systems. The concept of selfsimilarity is
proposed as an appropriate generalization of the well-known extensive
conditions exhibited by the traditional systems, which is used to justify a
given generalized thermodynamic formalism starting from the consideration of
the microcanonical ensemble, i.e. the nonextensive Statistics of Tsallis.
These ideas are illustrated by considering a recent proposed astrophysical
model based on the quasi-ergodic character of the microscopic dynamics of
these paradigmatic examples of real long-range interacting systems.

\end{abstract}
\date{\today}
\maketitle

\section{Introduction}

It is well-known that the divergence of the macroscopic response functions at
the critical point of the ferromagnetic-paramagnetic second order phase
transition is associated at the microscopic level with the divergence of the
correlation length $\xi$ among the systems constituents, phenomenon
interpreted as the establishment of a \textit{long-range order}. At the
dynamical framework, the divergence of $\xi$ also provokes a weakening of the
equilibration mechanisms (often diffusion), leading to a divergence of the
relaxation time $\tau_{r}$ when $T$ tends to the critical temperature $T_{c}$.
The characterization of the system properties in the neighborhood of the phase
transition has demanded an improvement of the Thermo-Statistical methods which
has leaded to the application of the Renormalization Group theory \cite{Gold}.
All that effort has been done to deal with the long-range correlations
existing at the critical point.

Let us now imagine a situation where all those points representing the
thermodynamic equilibrium states of a given system are dominated by the
existence of long-range correlations which are similar to the one discussed in
the above example. Obviously, the study of the thermodynamic quantities in
this context should be dominated by the presence of many anomalous behaviors
from the traditional point of view, representing important difficulties for
the classical formulations of the Thermodynamics and the Statistical
Mechanics. This last possibility is not a fictitious situation, contrary, this
is a feature of the called \textit{non extensive systems}.

The non extensive systems are those non necessarily composed by a huge number
of constituents, they could be mesoscopic or even small systems, where the
characteristic radio of some underlaying interaction is comparable or larger
than the characteristic linear dimension of the system, particularity leading
to the non existence of statistical independence due to the presence of
long-range correlations. In the last years a significant volume of evidences
of thermodynamic anomalous behaviors are been found in the study of the non
screened plasmas and the turbulent diffusion; astrophysical systems; nuclear,
molecular and atomic clusters; granular matter and complex systems
\cite{bog,bec,sol,shl,lyn,pie,pos,kon,tor,gro1,ato,kud,par,stil,sta1}. How
different are the thermodynamical properties of the non extensive systems can
be understood throughout the following examples.

According to the classical point of view, phase transitions only appear in the
thermodynamic limit \cite{yang-lee}. However, recent studies have shown the
existence of authentic phase transitions in small and mesoscopic systems. A
paradigmatic example is the nuclear multifragmentation, phenomenon observed
during the peripheral collisions of heavy ions, where a hundred of nucleons is
involved, which is interpreted, to the light of the new developments, as a
first order phase transition \cite{moretto, Dagostino}.

It is usual to find in this context equilibrium thermodynamic states which are
characterized by the presence of a \textit{negative heat capacity}, that is,
thermodynamic states where an increment of the total energy leads to a
decreasing of the temperature. This behavior is associated to the convexity of
the entropy and the corresponding \textit{ensemble inequivalence}. This
anomalous behavior could appear as a consequence of the small character of the
systems, i.e. during the first order phase transition in molecular and atomic
clusters \cite{gro na}, or because of the long-range character of the
interactions, i.e. the astrophysical systems \cite{antonovb}.

The presence of long-range correlations also leads to the existence of
relaxation mechanisms different from the ones operating in the traditional
systems. A paradigmatic example is the called \textit{violent relaxation},
very important to understand the evolution and structuration of the
astrophysical systems \cite{Lynden}. This mechanism is a\ very fast
equilibration process whose final state depends on the initial conditions, and
although it does not coincide with the Boltzmann-Gibss equilibrium, its
relaxation time $\tau_{r}$ is very large and in many cases diverges with the
imposition of the thermodynamic limit. The divergence of the relaxation time
provokes the non commutativity of the thermodynamic limit $N\rightarrow\infty$
and the infinite time limit $t\rightarrow\infty$ necessary for the
equilibration of the systems, and this fact is particularly important when
metastable states are present \cite{lat1,lat2,zanette,dauxois}.

The influence of some external conditions in the extensive systems, i.e. the
form of the container, only appears as a boundary effect, which do not change
the bulk properties. Contrary, the existence of long-range correlations in the
nonextensive systems leads to a strong dependence of the thermodynamics
properties on every external influence. This dependence of the thermodynamic
description on the external conditions could be dramatic in some cases, such
as the astrophysical systems. A selfgravitating system at a given energetic
state with a negative heat capacity can be in a stable thermodynamic state
whenever the system be isolated (microcanonical description). However, the
contact of the system with a heat bath at the same energetic state provokes
its gravitational collapse \cite{antonovb} as a consequence of the ensemble inequivalence.

The study of the nonextensive systems constitutes an interesting and
fascinating challenge for the developments of the Thermodynamics and
Statistical Mechanics. The purpose of the present study is to discuss some
aspects about the foundations of the macroscopic description of the
Hamiltonian systems, which are very important in physical applications and
where the classical formulations of the Thermo-statistics were established.
This choice also obeys to the available detailed knowledge about the features
of their microscopic dynamics, where we pay a special attention to the study
of the astrophysical systems, paradigmatic examples of the real non extensive systems.

\section{From microphysics to macrophysics}

The statistical entropy is closely related with the justification of a
thermodynamic formalism according to the viewpoint of Jaynes about the
reformulation of the Thermodynamics in terms of the Information Theory of
Shannon \cite{jaynes,shannon,kinchin}. However, the entropy as a measure of
the information is a vague concept, character leading to its nonuniqueness.
The clear manifestation of this character is found in the plethora of entropic
forms appearing in the scientific literature with different purposes, among
them, the generalization of the thermodynamic formalism
\cite{tsal,len,nau,kan,turc}. The interest is to derive an appropriate
macroscopic description\ for the non extensive Hamiltonian systems starting
from the general characteristics of their microscopic dynamics. This aim could
be mainly carried out by considering two\ alternative frameworks.

The first framework is the study of the nonequilibrium processes by
considering certain macroscopic dynamical equations. A central question is how
to derive some appropriate irreversible dynamical equations from the exact
microscopic reversible dynamics (i.e. the Liuville dynamics \cite{golstein})
and the justification of a given entropic form by considering a generalization
of the Boltzmann \textit{H}-theorem \cite{gallavotti,ajiezer}. Although this
is a very difficult way, this attempt is the most important because of it is
related with the justification of the Second Principle of Thermodynamics.

The second one is the study of the processes in thermodynamic equilibrium. The
equilibrium thermo-statistical description of the nonlinear Hamiltonian
systems with many degrees of freedom can be performed by considering the
microcanonical ensemble \cite{gallavotti}, whose justification relies on the
ergodic properties of the microscopic\ dynamics
\cite{Kolmogorov,Arnold,Moser,wiggins,PF1,PF2}. As elsewhere discussed, the
connection of this ensemble with the thermodynamic formalism is derived from
the general properties of the macroscopic observables, i.e. the well-known
extensive conditions of the traditional systems \cite{gallavotti}. In this
work we will concentrate our analysis in this kind of framework.

\section{Microscopic foundations\label{micfund}}

The impredictability associated with the dynamical sensibility under small
changes in the initial conditions, the called Hamiltonian Chaos, is considered
a necessary ingredient in the justification of the celebrate \textit{Ergodic
Hypothesis} which supports the tendency towards the equilibrium in most of
nonintegrable systems with many degrees of freedom and justifies the
applicability of the microcanonical ensemble in the Hamiltonian context
\cite{Kolmogorov,Arnold,Moser,wiggins,PF1,PF2}.%

\begin{figure}
[tb]
\begin{center}
\includegraphics[
height=2.8132in,
width=3.039in
]%
{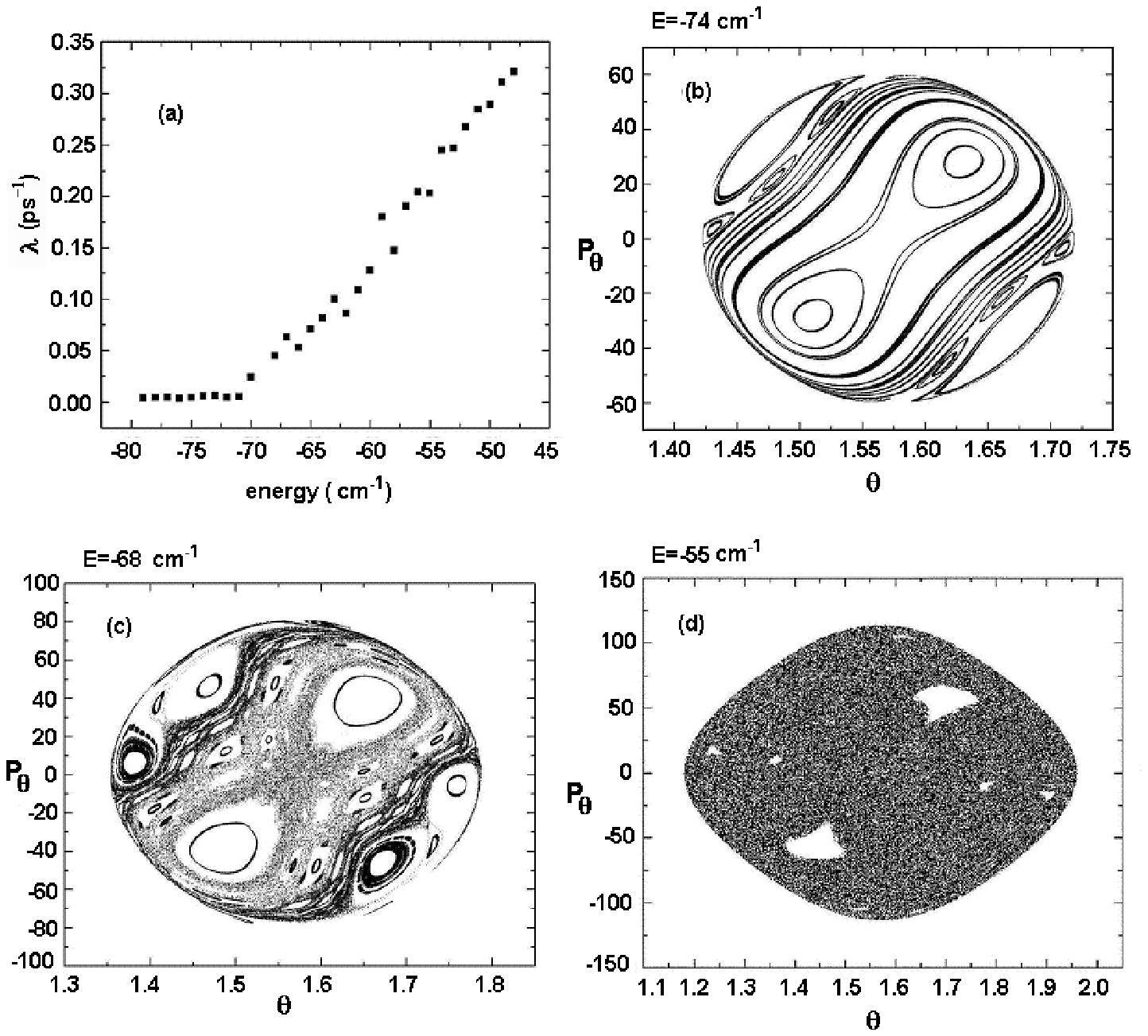}%
\caption{Relation between ergodicity and chaoticity in the $Ne..I_{2}$
molecule: Lyapunov exponents and the Poincare surface sections for different
values of the total energy.}%
\label{chaos}%
\end{center}
\end{figure}

This is clearly illustrated in systems with low dimensionality such as the
$Ne..I_{2}$ molecular complex \cite{vel3}\ shown in FIG.\ref{chaos}. During
the energy increasing of the system, the Lyapunov exponent shows a transition
from a stable dynamics towards a unstable one, while simultaneously the
Poincare Surface Sections show a transition from regular motions towards
irregular ones, where an arbitrary trajectory spreads uniformly in the whole
accessible phase space for a given energy.

Although the sufficient conditions for the existence of an irreversible
behavior justifying a statistical description of the macroscopic observables
have been mathematically shown only for some abstract models, there is a
general consensus that from the physical viewpoint most of the nonlinear
dynamical systems with many degrees of freedom are chaotic enough for
justifying a relaxation dynamics at least towards a quasi-stationary stage
\cite{gallavotti,lieberman,arnold}. \textquestiondown What happen when we move
to the context of the non extensive Hamiltonian systems? That is,
\textquestiondown Which are the microscopic mechanisms relevant in the case of
the long-range interacting \ systems leading to the chaoticity of the
Hamiltonian dynamics and allow us to justify a thermostatistical description
in this context?

The answers of these questions could be found by considering the recently
proposed Geometrical formalism of the Hamiltonian Chaos \cite{pettini4793}%
\ based on the well-known reformulation of the Hamiltonian dynamics in terms
of the Riemannian Geometry \cite{trieste}. As elsewhere discussed, in this
methodology the stability or instability of \ the system trajectories depends
on the curvature properties of some suitable defined manifold, feature
allowing us to unify the explanation about the origin of chaos with the most
extended quantity for its characterization: the Lyapunov exponents.

According to this formalism, the fundamental tool for describing the dynamical
sensibility is the called Jacobi-Levi-Civita equation \cite{Levi-Civita},
which adopts a very simple form for the special case of systems with two
degrees of freedom:%

\begin{equation}
\frac{d^{2}\xi}{ds^{2}}+\frac{1}{2}\mathcal{R}\left[  q\left(  s\right)
\right]  \xi=0,
\end{equation}
being $s$ the length of the trajectory $q\left(  s\right)  $ in the manifold,
$\mathcal{R}$, the scalar curvature of the configurational space, and $\xi$, a
quantity describing the spreading of two infinitely close trajectories. This
last equation allows us to understand in a simple fashion the origin of the
Hamiltonian chaos: Chaoticity can appear as a local effect due to the
existence of a negative curvature $\mathcal{R}$ in a given region of the
configurational space, or as a dynamical effect associated to the periodic or
quasi-periodic character of the system motions in bound configurations
whenever the curvature $\mathcal{R}$ be positive, the called mechanism of
\textit{parametric resonance} \cite{landau}.

The results obtained by using this methodology in different contexts suggest
us that the parametric resonance is possibly the most extended mechanism to
induce the chaoticity in systems with bound motions in the configurational
space \cite{pettini4793,vel3}. The incidence of the parametric resonance does
not depend on the short-range or long-range character of the underlying
interactions. This is the reason why this mechanism should be also relevant in
the case of the non extensive Hamiltonian systems, which have been
corroborated in recent studies \cite{pettini 51,pet}.

Taking into account the general consensus discussed above, the universality of
the microscopic mechanism of chaoticity shows that the long-range interacting
Hamiltonian systems satisfy those necessary condition justifying a
thermo-statistical description in the same fashion of the traditional systems,
suggesting us the\ great relevance of a thermo-statistical description with
microcanonical basis in this context. Such universality is reinforced by
considering all those conclusions and ideas derived from the assumption of the
\textit{Chaotic and Topological Hypotheses}. The first hypothesis puts on the
same level, for what concerns the statistical description of macroscopic
properties, chaotic many degrees of freedom dynamical systems and
\textit{Anosov systems} \footnote{Among the abstract models, the Anosov
systems are those characterized by the strongest statistical properties.}
\cite{gallavotti,cohenG}, and the second one establishes that both the origin
of the chaoticity of the microscopic dynamics and the phase transitions at the
macroscopic level depend on the geometrical and topological structure of the
configurational space \cite{topH2}.

\section{Macroscopic Foundations}

The microcanonical ensemble is properly a \textit{dynamical ensemble}, where
all the macroscopic observables derived from it have a direct mechanical
interpretation. This is the reason why this ensemble represents a direct
connection between the microscopic and macroscopic descriptions of the
Hamiltonian systems in thermodynamic equilibrium, constituting a solid way in
the thermo-statistical characterization of these systems without considering
anything outside the Mechanics.

This is precisely the viewpoint of Boltzmann, Gibbs, Einstein and Ehrenfest,
who showed the hierarchical primacy of the microcanonical ensemble in regard
with the Canonical and Gran canonical ensembles \cite{Bolt,Gibb,Ein,Ehre},
which can be derived from the first one by considering the extensive
conditions \cite{pat,stan,land}. As elsewhere discussed, the extensive
conditions are implemented by taking into consideration the well-known Gibbs
argument about a subsystem with a weakly interaction with a heat bath, or the
ensemble equivalence in the thermodynamic limit \cite{gallavotti}.

\textquestiondown Could this extensive conditions be generalized in order to
allow a natural extension of the traditional formalism starting from
microcanonical basis? The affirmative answer to this question leads to the
\textit{selfsimilarity concept} and the notion about a \textit{generalized
canonical ensemble}. Since the separability in subsystems can not be ensured
for non extensive systems due to the long-range correlations existing in this
context, the Gibbs argument is extremely limited in order to allowing such a
generalization. Thus, the ensemble equivalence in the thermodynamic limit has
a more general applicability in deriving an appropriate thermo-statistical
description of the Hamiltonian non extensive systems.

\subsection{Selfsimilarity scaling laws}

The extensive character of the traditional systems is manifested during the
scaling transformation of the system size $N$ in $\alpha$ times,
$N\rightarrow\alpha N$, operation leading to a similar scaling behavior of
certain fundamental set of macroscopic observables such as the total energy
$E$, the system volume $V$, etc. This physical quantities are ordinarily
referred as \textit{extensive observables}. This scaling operation leads to an
exponential growing of the number of microscopic states (or the accessible
phase space volume) $W=W\left(  E,N,V\right)  $ of the form:%

\begin{equation}
\left.
\begin{array}
[c]{c}%
N\left(  \alpha\right)  =\alpha N\\
E\left(  \alpha\right)  =\alpha E\\
V\left(  \alpha\right)  =\alpha V
\end{array}
\right\}  \Rightarrow W\left[  \alpha\right]  =W^{\alpha},
\end{equation}
mathematical property leading to the extensive character of the Boltzmann
entropy $S_{B}=\ln W$:%

\begin{equation}
W^{\alpha}=\exp\left[  \alpha\ln W\right]  \Rightarrow S_{B}\left(
\alpha\right)  =\alpha S_{B}. \label{bolt_ent}%
\end{equation}
Of course, the extensivity appears asyntotically with the imposition of the
thermodynamic limit:%
\begin{equation}
N\text{ tends to infinity, keeping }\frac{E}{N}\text{ and }\frac{N}{V}\text{
fixed.}%
\end{equation}

A natural extension of extensivity leads to the consideration of certain
multiplicative uniparametric group $T_{\alpha}$, $T_{\alpha}T_{\beta
}=T_{\alpha\beta}$, acting on those fundamental macroscopic observables $I$
determining the microcanonical macroscopic state, which leads to a scaling
transformation $F_{\alpha}$ of the number of microscopic configurations
$W=W\left(  I\right)  $ whose form does not depend on the macroscopic state:%

\begin{equation}
I\left(  \alpha\right)  =T_{\alpha}I\Rightarrow W\left[  \alpha\right]
=F_{\alpha}\left(  W\right)  .
\end{equation}
We say that a given system exhibits a \textit{scaling self-similarity} when
this kind of symmetry appears during the scaling operation.

A rigorous mathematical definition of selfsimilarity can be established as
follows. Let us consider a n-dimensional Euclidean space $\mathcal{M}$ and a
map $W$ $:\mathcal{M}\rightarrow\mathcal{R}^{+}$ relating every point
$I\in\mathcal{M}$ with a positive real number $w=W\left(  I\right)
\in\mathcal{R}^{+}$. It is said that the map $W$ exhibits a scaling
selfsimilarity if \ there is a uniparametric subgroup $\wp_{\mathcal{M}}$
belonging to the group of the automorphism of the space\textit{ }$\mathcal{M}%
$:\textit{ }%
\begin{equation}
\wp_{\mathcal{M}}=\left\{  T_{\alpha}:\mathcal{M}\rightarrow\mathcal{M}%
\right\vert \left.  \alpha\in\mathcal{R}^{+}~\right\}  ,
\end{equation}
satisfying the following properties:

\begin{description}
\item[ \textbf{\ i)}] $\wp_{\mathcal{M}}$ is a multiplicative\textit{ group},
that is, $\forall\alpha,\beta\in\mathcal{R}^{+}$ $\Rightarrow T_{\alpha
}T_{\beta}=T_{\alpha\beta}$.

\item[ \textbf{ii)}] $\forall~T_{\alpha}\in\wp_{\mathcal{M}}$ and $\forall$
$I\in\mathcal{M}$, there exists a positive real function $F_{\alpha
}:\mathcal{R}^{+}\rightarrow\mathcal{R}^{+}$ obeying the relation $W\left[
T_{\alpha}\left(  I\right)  \right]  =F_{\alpha}\left[  W\left(  I\right)
\right]  $.

\item[\textbf{iii)}] $\forall x,y\in\mathcal{R}^{+}$ there is only one
function $F_{\alpha}\in\wp_{\mathcal{R}}$ satisfying the relation
$x=F_{\alpha}\left(  y\right)  $, being $\wp_{\mathcal{R}}$ the set composed
by all the functions $F_{\alpha}$ above referred:%
\begin{equation}
\wp_{\mathcal{R}}=\left\{  F_{\alpha}:\mathcal{R}^{+}\rightarrow
\mathcal{R}^{+}\right\vert \left.  \alpha\in\mathcal{R}^{+}~\right\}  .
\end{equation}

\end{description}

Now on $\wp_{\mathcal{M}}$ will be referred as the\textit{ group} of
the\textit{ scaling} \textit{selfsimilarity} transformations of the space
$\mathcal{M}$, being $\alpha$ the scaling parameter. The condition
\textbf{ii)} establishes an isomorphic correspondence $\mathcal{K}%
:\wp_{\mathcal{M}}\rightarrow$ $\wp_{\mathcal{R}}$ among the elements
$T_{\alpha}\in\wp_{\mathcal{M}}$ and the functions $F_{\alpha}\in
\wp_{\mathcal{R}}$. Properties \textbf{ii)} and \textbf{iii)} allow us to say
that $\wp_{\mathcal{R}}$ is a subgroup of the group of automorphism of
$\mathcal{R}^{+}$.

Let us consider two arbitrary transformations $T_{\alpha},T_{\beta}\in
\wp_{\mathcal{M}}$, the transformation $T_{\alpha\beta}=T_{a}T_{\beta}$, as
well as their corresponding isomorphic applications $F_{\alpha}$,$F_{\beta}$
and $F_{\alpha\beta}$ in the group $\wp_{\mathcal{R}}$. The following
transformations are straightforwardly derived from the property \textbf{ii)}:%

\begin{align}
W\left[  T_{\alpha}T_{\beta}\left(  I\right)  \right]   &  =W\left[
T_{\alpha}\left(  T_{\beta}I\right)  \right]  =F_{\alpha}\left[  W\left(
T_{\beta}I\right)  \right] \nonumber\\
&  =F_{\alpha}\left[  F_{\beta}W\left(  I\right)  \right]  =F_{\alpha}%
F_{\beta}\left[  W\left(  I\right)  \right]  ,
\end{align}
Taking into account the relation:
\begin{equation}
W\left[  T_{\alpha}T_{\beta}\left(  I\right)  \right]  =W\left(
T_{\alpha\beta}\left(  I\right)  \right)  =F_{\alpha\beta}\left[  W\left(
I\right)  \right]  ,
\end{equation}
we obtain the following result:%

\begin{equation}
F_{\alpha}F_{\beta}=F_{\alpha\beta}.
\end{equation}
Therefore, the correspondence $\mathcal{K}$ defines a homomorphism between the
groups $\wp_{\mathcal{M}}\ $and $\wp_{\mathcal{R}}$.

Let us now consider an arbitrary point $x_{0}\in\mathcal{R}^{+}$ and denote by
$\mathcal{Y}_{0}\left(  \alpha\right)  $ its image point $\mathcal{Y}%
_{0}\left(  \alpha\right)  =F_{\alpha}\left(  x_{0}\right)  $ associated with
the scaling transformation $T_{\alpha}$. The property \textbf{iii)} allows us
to say that the function $y=\mathcal{Y}_{0}\left(  \alpha\right)  $ is
invertible in regard with the scaling parameter $\alpha$ because of there is
only one scaling transformation relating the points $y$ and $x_{0}$. Let
$\mathcal{Y}_{0}^{-1}\left[  y\right]  $ be the inverse of the function
$\mathcal{Y}_{0}\left(  \alpha\right)  $, $\alpha=\mathcal{Y}_{0}^{-1}\left[
\mathcal{Y}_{0}\left(  \alpha\right)  \right]  $. Taking into account that
$F_{\alpha\beta}\left(  x_{0}\right)  =\mathcal{Y}_{0}\left(  \alpha
\beta\right)  =F_{\alpha}\left[  F_{\beta}\left(  x_{0}\right)  \right]
=F_{\alpha}\left[  \mathcal{Y}_{0}\left(  \beta\right)  \right]  $, and
substituting the identity $\beta=\mathcal{Y}_{0}^{-1}\left[  \mathcal{Y}%
_{0}\left(  \beta\right)  \right]  $, we obtain the relation $F_{\alpha
}\left[  \mathcal{Y}_{0}\left(  \beta\right)  \right]  =\mathcal{Y}%
_{0}\left\langle \alpha\mathcal{Y}_{0}^{-1}\left[  \mathcal{Y}_{0}\left(
\beta\right)  \right]  \right\rangle $. Since $\beta$ is an arbitrary positive
real number, we set $x=\mathcal{Y}_{0}\left(  \beta\right)  $ and rewrite the
last relation in the form:%

\begin{equation}
F_{\alpha}\left(  x\right)  =\mathcal{Y}_{0}\left[  \alpha\mathcal{Y}_{0}%
^{-1}\left(  x\right)  \right]  . \label{descomposicion}%
\end{equation}
Obviously, the function $F_{\alpha}\left(  x\right)  $ does not depend on the
selection of the point $x_{0}$ used for defining the function $\mathcal{Y}%
_{0}\left(  \alpha\right)  $. Considering another point $x_{1}$ and denoting
by $\mathcal{Y}_{1}\left(  \alpha\right)  $ the function $\mathcal{Y}%
_{1}\left(  \alpha\right)  =F_{\alpha}\left(  x_{1}\right)  $, the property
\textbf{iii)} ensures the existence of a unique positive real number $\beta$
satisfying the relation $x_{1}=F_{\beta}\left(  x_{0}\right)  =\mathcal{Y}%
_{0}\left(  \beta\right)  $, so that $\beta=\mathcal{Y}_{0}^{-1}\left(
x_{1}\right)  $. Substituting $x$ by $x_{1}$ in the equation
(\ref{descomposicion}):%

\begin{align}
F_{\alpha}\left(  x_{1}\right)   &  =\mathcal{Y}_{0}\left[  \alpha
\mathcal{Y}_{0}^{-1}\left(  x_{1}\right)  \right] \nonumber\\
&  =\mathcal{Y}_{0}\left(  \alpha\beta\right)  =\mathcal{Y}_{1}\left(
\alpha\right)  ,
\end{align}
and applying the inverse function $\mathcal{Y}_{0}^{-1}\left(  x\right)  $,
the relation $\mathcal{Y}_{0}^{-1}o\mathcal{Y}_{1}\equiv\beta\mathcal{I}$ is
obtained, being $\mathcal{I}$ the identity function. Analogue reasonings lead
to the relation $\mathcal{Y}_{1}^{-1}o\mathcal{Y}_{0}\equiv\beta
^{-1}\mathcal{I}$, and hence:
\begin{equation}
\mathcal{Y}_{1}\left(  x\right)  =\beta\mathcal{Y}_{0}\left(  x\right)  .
\end{equation}
Therefore, the equation (\ref{descomposicion}) can be rewritten as:%

\begin{equation}
F_{\alpha}\left(  x\right)  =\mathcal{Y}\left[  \alpha\mathcal{Y}^{-1}\left(
x\right)  \right]  , \label{scaling}%
\end{equation}
where the function $\mathcal{Y}\left(  x\right)  $ is determined till the
precision of a constant factor.

It has been shown in this way that the selfsimilarity scaling transformation
$F_{\alpha}\left(  W\right)  $ could be rewritten by using certain function
$\mathcal{Y}\left(  x\right)  $ in an analogue fashion that the exponential
function in the equation (\ref{bolt_ent}) considers the exponential growing of
the number of microscopic configurations $W$ in the extensive systems. We say
that the function $\mathcal{Y}\left(  x\right)  $\ determines the
\textit{selfsimilarity scaling law} of the system.

\subsection{Generalized Boltzmann Principle}

The Boltzmann entropy is derived from the Shannon-Boltzmann-Gibbs extensive
entropy \cite{jaynes,shannon,kinchin}:%

\begin{equation}
S_{BG}=-\sum_{k}p_{k}\ln p_{k}, \label{sbg}%
\end{equation}
by considering the equalprobability property which characterizes the
microcanonical description. The logarithmic dependence is essential to support
the additivity of the entropy under the consideration of a close system
composed by two independent subsystems $A$ and $B$:%

\begin{equation}
S\left(  A\oplus B\right)  =S\left(  A\right)  +S\left(  B\right)  .
\label{aditividad}%
\end{equation}

According to the Kinchin postulates \cite{kinchin}, the extensive entropy is
the only one entropic form satisfying simultaneously the additivity and
concavity conditions. Many authors think that the additivity of the extensive
entropy is the origin of its non applicability to the non extensive systems,
being this idea the main motivation for the consideration of many others
entropic forms \cite{tsal,len,nau,kan,turc} by generalizing the fourth Kinchin
postulate \cite{kinchin}.

Outside the Information Theory context, the logarithmic dependence of the
extensive entropy is related with the wide variety of behaviors obeying
exponential laws. Among them, there are two very relevant: (1) the exponential
sensibility of the dynamical trajectories and (2) the exponential growing of
the accessible volume $W$\ with the increasing of the system size.

As elsewhere discussed, the dynamical sensibility can be quantified by using
the Lyapunov exponents. This last quantity is intimately related with another
measure used for characterizing the chaos based on the consideration of the
Shannon-Boltzmann-Gibbs extensive entropy: the\textit{ Komolgorov-Sinai
entropy}. According to the theory of dynamical systems, the Komolgorov-Sinai
entropy is numerically equal to the average value of the positive Lyapunov
exponents on the attractor points \cite{hilborn,pesin}, evidence suggesting a
strong connection between the character of the dynamical sensibility and the
entropic form relevant at the macroscopic scale. This last possibility has
been recently used in order to justify a generalized entropic form by
considering a non exponential sensibility of the microscopic dynamics
\cite{robledo,rob2,coradu,baldovin,wada,latorab}.

As already shown, the exponential growing of the accessible volume $W$ during
the scaling transformation leads to the extensivity of the Boltzmann entropy,
being this an essential condition in order to ensure the equivalence between
the microcanonical ensemble and the Canonical or Gran canonical ensembles
during the imposition of the thermodynamic limit. This ensemble equivalence
supports the Legendre transformation among the thermodynamic potentials in
which is based the thermodynamic formalism \cite{gallavotti}. Our opinion is
that the logarithmic form of the Boltzmann entropy can not be an \textit{ad
hoc} definition. Contrary, such logarithmic form of the Boltzmann entropy is
\textit{determined from} the exponential growing of the accessible volume $W$
during the scaling transformations.

Taking into consideration this last idea, the existence of selfsimilarity in a
given non extensive system should lead to assume a generalized form for the
Boltzmann entropy compatible with the underlaying selfsimilarity scaling law.
The analogy between the equations (\ref{bolt_ent}) and (\ref{scaling})
suggests us the consideration of the following \textit{generalized Boltzmann
Principle}:%
\begin{equation}
S_{B}^{\mathcal{L}}=\mathcal{L}\left(  W\right)  , \label{g_bolt}%
\end{equation}
where $\mathcal{L}\left(  x\right)  $ is the inverse of the function
$\mathcal{Y}\left(  x\right)  $ which determines the selfsimilarity scaling
law of the system, $\mathcal{L}\left(  x\right)  =\mathcal{Y}^{-1}\left(
x\right)  $.

Assuming a generic statistical entropy with the following trace form:%
\begin{equation}
S_{h}\left[  p\right]  =\sum_{k}h\left(  p_{k}\right)  ,
\end{equation}
and considering a given system with a number $W$ of equiprobable microscopic
states (a microcanonical description with $p_{k}=1/W$), the identification of
this microcanonical entropy with the generalized Boltzmann entropy
(\ref{g_bolt}):
\begin{equation}
S_{h}\left[  p_{mic}\right]  =W\ast h\left(  \frac{1}{W}\right)
=\mathcal{L}\left(  W\right)  ,
\end{equation}
leads to assume the following \textit{generalized statistical entropy}
compatible with the selfsimilarity of a given system:%
\begin{equation}
S_{e}^{\mathcal{L}}=\sum_{k}p_{k}\mathcal{L}\left(  \frac{1}{p_{k}}\right)  .
\label{gse}%
\end{equation}

This last statistical entropy should be considered as the starting point for
the derivation of a generalized Canonical ensemble and the generalization of
the thermodynamic formalism. As already discussed, the thermodynamic formalism
should be derived from the analysis of the conditions for the ensemble
equivalence in the thermodynamic limit. Such thermodynamic limit can not be
arbitrarily introduced, contrary, it should be compatible with the underlying
scaling selfsimilarity.

\subsection{Some generalized Statistics}

Most of the Hamiltonian systems with a practical interest should exhibit an
\textit{exponential }growing of the accessible phase space volume $W$ with the
increasing of the system degrees of freedom. This result could be easily shown
by considering a generic Hamiltonian system of the form:%
\begin{equation}
H\left(  q,p\right)  =\sum_{ij}\frac{1}{2}a^{ij}\left(  q\right)  p_{i}%
p_{j}+V\left(  q\right)  , \label{generic}%
\end{equation}
where $i=1,2,...n$, and $q=\left(  q^{1},q^{2},\ldots q^{n}\right)  $
represents the coordinates in the configurational space. The microcanonical
accessible phase space volume $W$:%
\begin{equation}
W=\delta\varepsilon\int\frac{d^{n}qd^{n}p}{h^{n}}\delta\left[  E-H\left(
q,p\right)  \right]  ,
\end{equation}
can be rewritten after the integration over the momenta as follows:
\[
W=\frac{\delta\varepsilon}{\Gamma\left(  n\right)  }\left(  \frac{2\pi}%
{h}\right)  ^{n}\int_{M}\frac{d\mu\left(  q\right)  }{E-V\left(  q\right)  },
\]
being $d\mu\left(  q\right)  =\sqrt{\left\vert g_{ij}\left(  q\right)
\right\vert }d^{n}q$ the volume element of the configurational space manifold
$M$ derived from the configurational space with the consideration of the
Jacobi metric:%

\begin{equation}
ds^{2}=g_{ij}\left(  q\right)  dq^{i}dq^{j}=2\left[  E-V\left(  q\right)
\right]  a_{ij}\left(  q\right)  dq^{i}dq^{j}, \label{metrica}%
\end{equation}
with $a_{ij}\left(  q\right)  =\left\langle a^{ij}\left(  q\right)
\right\rangle ^{-1}$. Since, $\sqrt{\left\vert g_{ij}\left(  q\right)
\right\vert }\propto\left[  E-V\left(  q\right)  \right]  ^{\frac{1}{2}n}$,
the increasing of the system degrees of freedom $n$ leads to an exponential
growing of $W$. Interestingly, the chaoticity of such Hamiltonian dynamics can
be described by using the Jacobi-Levi-Civita equation of the Geometrical
formalism \cite{pettini4793}, whose linear character leads also to the
\textit{exponential} character of the dynamical sensibility of the system trajectories.

According to our selfsimilarity viewpoint, the above observations suggest
strongly that the classical Thermo-statistics is still applicable to many non
extensive systems, that is, the existence of long-range correlations due to
the presence of long-range interactions in the potential energy $V\left(
q\right)  $ is not a sufficient condition for invaliding the applicability of
the Shanonn-Boltzmann-Gibbs extensive entropy in this context. Of course, some
additional requirements are demanded in order to consider the non extensivity
of the energy and other integrals of motions determining the macroscopic
state. The complete description of the specific exponential selfsimilarity
obeyed by a given system demands also the determination of the scaling
transformation group $T_{\alpha}$ acting on the fundamental macroscopic
observables, question not considered in this paper.

Since not all Hamiltonian systems should obey the generic form (\ref{generic}%
), not all Hamiltonian systems should necessarily satisfy exponential
selfsimilarity scaling laws. We have shown in ref.\cite{velTsal} that the
consideration of a Hamiltonian system with an underlying potential scaling law
should be described by the following \textit{q}-generalized Boltzmann entropy:%
\begin{equation}
\left(  S_{B}\right)  _{q}=\ln_{q}W=\frac{W^{1-q}-1}{1-q},
\end{equation}
associated with the popular \textit{Tsallis nonextensive entropy} \cite{tsal}:%
\begin{align}
S_{q}  &  =\sum_{k}p_{k}\ln_{q}\left(  \frac{1}{p_{k}}\right)  \equiv-\sum
_{k}p_{k}^{q}\ln_{q}p_{k},\nonumber\\
&  =\left(  q-1\right)  ^{-1}\sum_{k}\left(  p_{k}-p_{k}^{q}\right)  .
\end{align}
We used here the q-logarithmic function $\ln_{q}x=e_{q}^{-1}\left(  x\right)
$, which satisfies the\ identity $\ln_{q}\left(  \frac{1}{x}\right)
=-x^{q-1}\ln_{q}x$. Such functions become in the ordinary exponential and
logarithmic functions when $q=1$.

It is remarkable that many results derived from the Tsallis nonextensive
Statistics should be obtained in a natural fashion from the selfsimilarity
ideas, i.e. the q-expectation values, the generalized Legendre transformation,
etc., where the enigmatic entropic index $q$ could be obtained from an
optimization procedure without considering anything outside the Mechanics
\cite{velTsal}. Unfortunately, we do not known any example of a Hamiltonian
system satisfying a potential selfsimilarity scaling law, so that, the
applicability of the Tsallis Statistics is still an open problem. There are
some experimental and theoretical studies suggesting that the Tsallis
formalism could be useful for describing nonequilibrium situations in
long-range interacting Hamiltonian systems in metastable states
\cite{hs1,hs3,hs4}. The entropic index $q$ have been related with a potential
sensibility of the microscopic dynamics in the onset of chaos
\cite{robledo,rob2,coradu,baldovin,wada,latorab} described by a generic
tangent dynamics of the form:
\begin{equation}
\frac{d\xi}{dt}=\lambda_{q}\xi^{q}\text{, }\xi\left(  0\right)  =1\Rightarrow
\xi\left(  t\right)  =e_{q}\left(  \lambda_{q}t\right)  ,
\end{equation}
recovering the ordinary exponential sensibility for $q=1$.

\section{Astrophysical systems}

An important applicability framework of the above ideas could be found in the
study of the astrophysical systems. As elsewhere discussed, the astrophysical
systems exhibit many similarities in their structural and the dynamical
behaviors which are independent from the scale where such processes take
place, i.e. Vacoulers and Sercic laws, the velocity distribution, violent
relaxation, gravitational collapse
\cite{hjorth,sersic,chandra,spitzer,michie,king,bin}. The existence of such
regularities could be only explained by the incidence of certain relaxation
mechanisms leading these systems to quasi-stationary configurations, which, in
principle, could be described by using a thermo-statistical description.

However, in spite of the Gravitation is the oldest known physical interaction,
it possesses very important singularities that do not allow us to perform a
rigorous thermo-statistical description of the astrophysical systems. We refer
to the well-known \textit{short-range} and \textit{long-range singularities}
of the Newtonian gravitational interaction.

The short-range singularity is the divergence of the gravitational potential
$\varphi\left(  r\right)  $ when $r\rightarrow0$ allowing to a given finite
total energy the existence of microscopic configurations with infinite
absolute values of the kinetic and potential total energies, which are
characterized by exhibiting a very dense core with central density $\rho
_{c}\rightarrow\infty$ and a diluted halo (collapsed configurations). The
long-range singularity is related with the long-range character of the
gravitational interaction: although $\varphi\left(  r\right)  $ drops to zero
when $r\rightarrow\infty$, the vanishing is very slow, feature leading to the
existence of long-range correlations among the constituent particles because
of the infinite value of the interaction radio.

These singularities have the following important consequences: (1) existence
of a gravitational collapse at low energies, (2) the non extensive character
of the astrophysical systems due to the infinite radio of the gravitational
interaction, (3) macroscopic configurations characterized by the presence of a
negative heat capacity which does not disappear with the imposition of the
thermodynamic limit, and the consequent ensemble inequivalence, (4) the
incidence of a particle evaporation and a incomplete relaxation of the system,
(5) the missing of a real thermodynamic equilibrium and the dynamical
ergodicity due to the non enclosed character of the\ system motion.%

\begin{figure}
[tb]
\begin{center}
\includegraphics[
height=2.7216in,
width=3.2396in
]%
{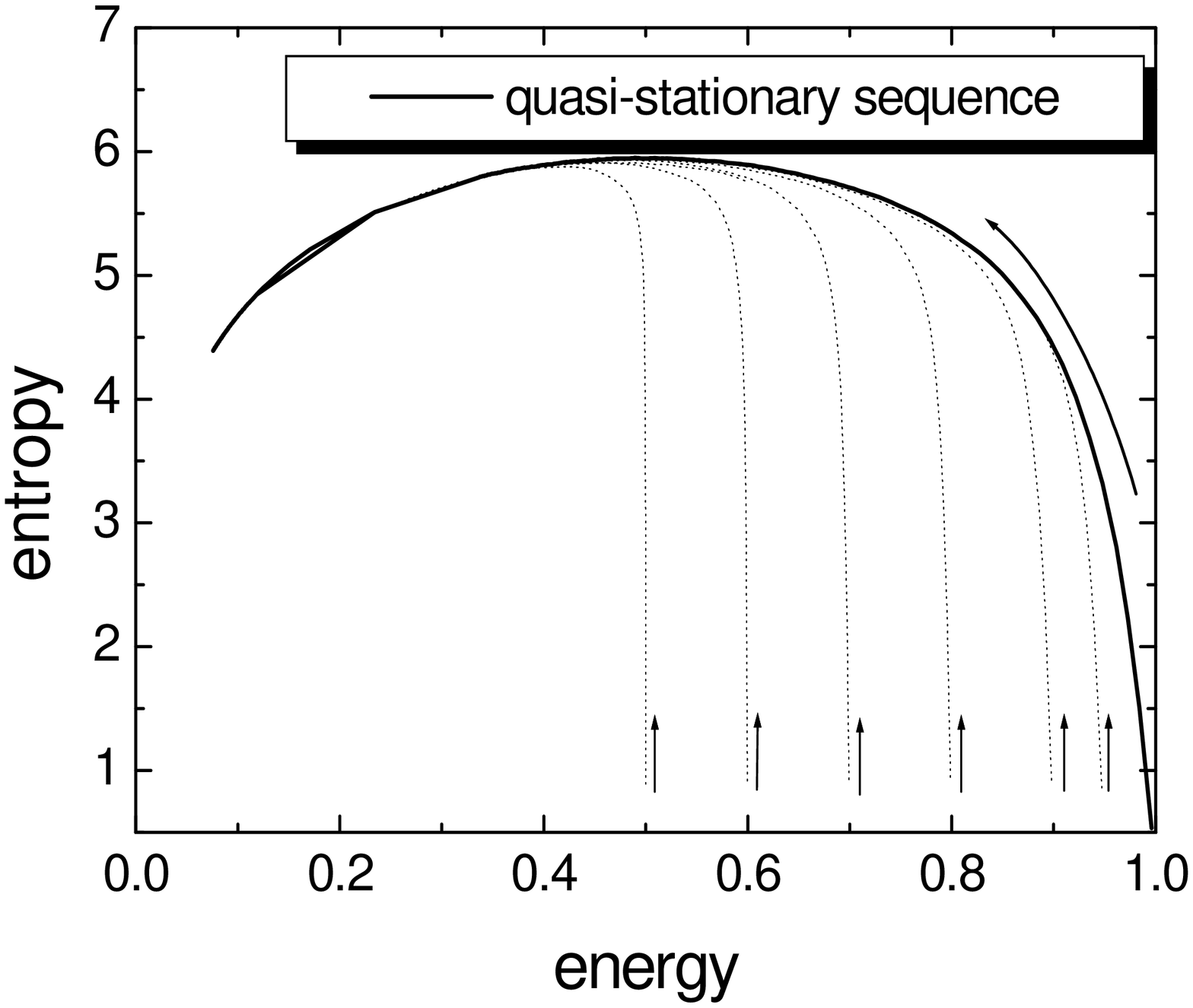}%
\caption{Quasi-stationary evolution in a\ gas of binary encounters driven by a
particle evaporation. }%
\label{sequence}%
\end{center}
\end{figure}

The application of a thermo-statistical description in this context demands a
close analysis of the underlying microscopic picture, where it is very
important to understand the role of the particle evaporation in the nature of
the macroscopic characterization of the astrophysical systems.
FIG.\ref{sequence} shows the entropy-energy diagram obtained from the
dynamical study \cite{velevap} of a modified version of the well-known
Spitzer-H\"{a}rm model used for justifying the Michie-King model of globular
clusters and elliptical galaxies \cite{spitzer,michie,king,bin}. We observe
here that every dynamical trajectory (dot lines) tries to follow certain
sequence (solid line) which is independent of the initial conditions. This
behavior accounts for the existence of a quasi-stationary evolution which is
established without mattering how intense is the evaporation, although it
depends on every detail of the microscopic dynamics \cite{velevap}.

Thus, the thermo-statistical description of the astrophysical system should
start from the consideration of the quasi-stationary character of the
evolution of these systems as well as the incidence of several relaxation
mechanism, such as the binary collisions \cite{bin} and the parametric
resonance \cite{pet} already commented in section \ref{micfund}. Recent
studies showed that the characteristic chaotization time $\tau_{ch}$ of the
dynamical trajectories of such systems is determined by the incidence of the
parametric resonance \cite{pettini 51,pet}, which talks about certain
predominance of this relaxation mechanism over the binary encounters:%

\begin{equation}
\tau_{ch}\sim\tau_{mic}<<\tau_{coll}\sim\tau_{mac},
\end{equation}
where $\tau_{mac}\sim0.1\tau_{mic}N/\ln N$ is the characteristic relaxation
time derived from the consideration of the binary collisions, $\tau
_{mic}=1/\sqrt{G\rho_{c}}$, the characteristic microscopic time, and $\rho
_{c}$, the characteristic density of the system. This results suggests us that
those astrophysical systems with a collisionless dynamical evolution are
chaotic enough for supporting a thermo-statistical description based on a
quasi-ergodicity of the microscopic dynamics \cite{pet}.%

\begin{figure}
[t]
\begin{center}
\includegraphics[
height=3.6348in,
width=3.039in
]%
{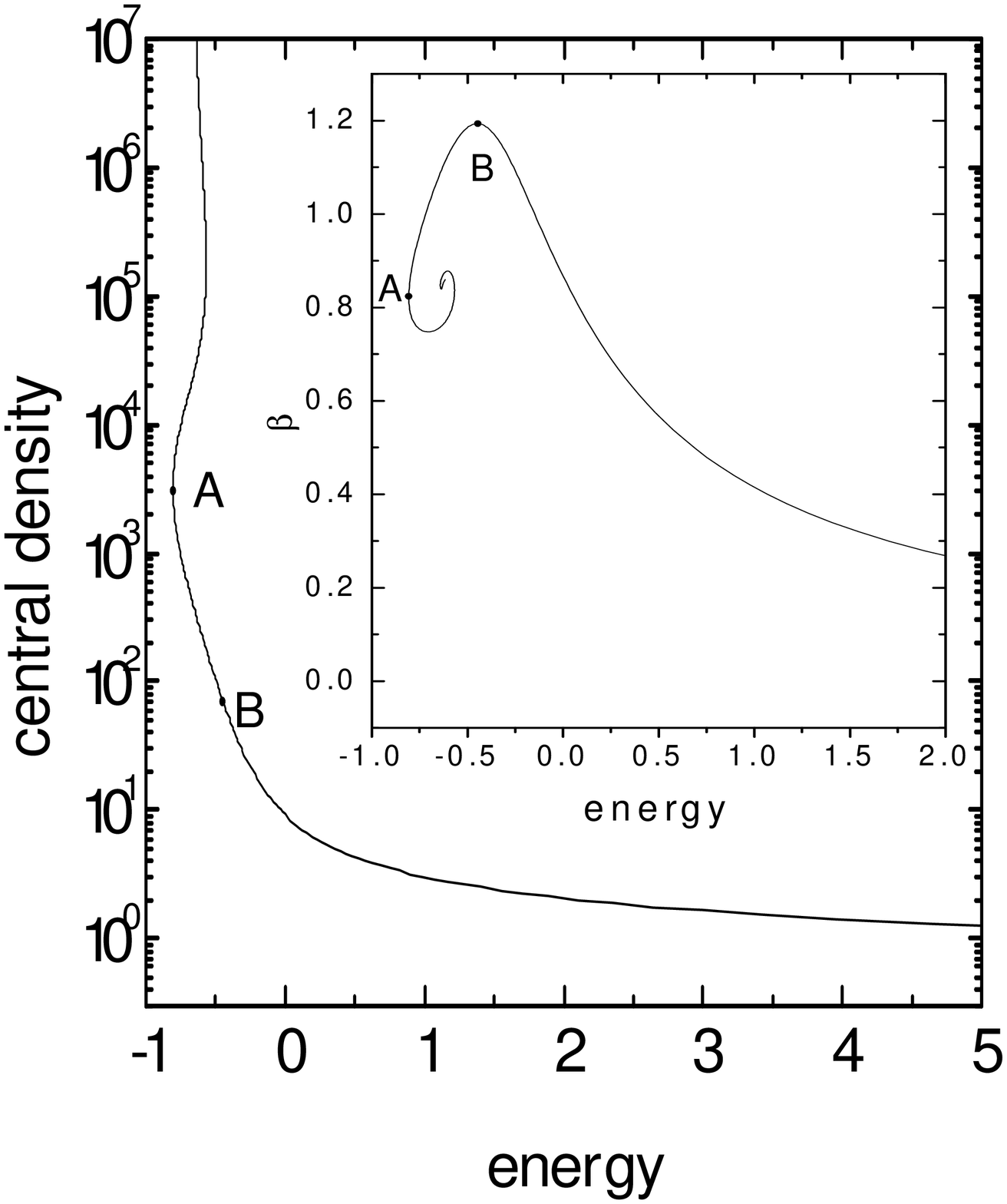}%
\caption{Thermo-statistical description of the alternative version of the
Antonov isothermal model: central density and inverse temperature (inserted
graph) \textit{versus} energy. }%
\label{alternativo}%
\end{center}
\end{figure}

The above macroscopic foundations were used in order to implement an
alternative version of the Antonov isothermal model \cite{velastro} which
keeps several similarities with the Michie-King model
\cite{chandra,spitzer,michie,king,bin}. Remarkably, this description performs
a realistic characterization of the phenomena taking place in this context by
using a purely thermodynamic language. FIG.\ref{alternativo} shows the
dependence of the central density \textit{versus} the total energy of the
system, as well as the caloric curve (inverse temperature $\beta$
\textit{versus} energy), which shows the existence of a gravitational collapse
and the persistence of thermodynamic states with negative head capacity at low energies.

We present here the specific selfsimilarity scaling laws exhibited by this
model system \cite{velastro}, relevant in the case of the gas of identical non
relativistic point particles interacting by means of the Newtonian gravity:%

\begin{equation}
\left.
\begin{array}
[c]{c}%
N\left(  \alpha\right)  =\alpha N\\
E\left(  \alpha\right)  =\alpha^{\frac{7}{3}}E\\
L\left(  \alpha\right)  =\alpha^{-\frac{1}{3}}L
\end{array}
\right\}  \Rightarrow W\left(  \alpha\right)  =\exp\left[  \alpha\ln W\right]
, \label{self}%
\end{equation}
where $L$ is the characteristic linear dimension of the system. Notice that
although this scaling transformation differs from the one acting on the
extensive systems, they exhibit the same exponential selfsimilarity scaling
law, and consequently, the selfsimilarity properties of the astrophysical
systems support the applicability of the Boltzmann-Gibbs statistics.
The\ relevant thermodynamic limit in this case is that remaining unchanged
under the scaling transformations (\ref{self}):%

\begin{equation}
\text{to tend }N\rightarrow\infty\text{, keeping fixed }\frac{E}{N^{\frac
{7}{3}}}\text{ and }LN^{\frac{1}{3}}. \label{STHL}%
\end{equation}

The above thermodynamic limit has been also obtained by other investigators
\cite{chava12}, although it differs from the one reported by other authors
\cite{de vega}. The essential difference among them is based on the fact that
these results were obtained by considering other suppositions, such as the
one\ obtained by using the well-known Kac criterion \cite{kac}:%

\begin{equation}
\text{to tend }N\rightarrow\infty\text{, keeping fixed }\frac{E}{N}\text{ and
}\frac{L}{N}, \label{KTHL}%
\end{equation}
where the model parameters are \textit{a priory} scaled by some power of the
system size $N$ in order to deal with an extensive total energy. According to
our opinion, the Kac criterion is an unphysical assumption because of,
although the extensive character of the total energy is forced, the system
remains essentially nonextensive since it could not be divided into
independent subsystem even in the thermodynamic limit. Besides, the scaling of
the system parameters \textit{affects} the characteristic temporal scale of
the evolution.

Taking into account the above estimation of the relaxation time derived from
the consideration of \ binary encounters:%

\begin{equation}
\tau_{rel}\sim\sqrt{\frac{L^{3}}{GM}}\frac{N}{\ln N},
\end{equation}
it can be verified that $\tau_{rel}$ remains finite when the thermodynamic
limit derived from the selfsimilarity (\ref{STHL}) is imposed (dismissing the
logarithmic factor with zero-order of growing), while it diverges when the one
associated with the Kac criterion (\ref{KTHL}) is considered . Such dynamical
anomaly implies the noncommutativity between the thermodynamic limit and the
infinite time limit necessary for the macroscopic equilibration:%

\begin{align}
\lim_{N\rightarrow\infty}\lim_{T\rightarrow\infty}\left\langle F_{N}%
\right\rangle _{T}  &  \not =\lim_{T\rightarrow\infty}\lim_{N\rightarrow
\infty}\left\langle F_{N}\right\rangle _{T},\nonumber\\
\text{where }\left\langle F_{N}\right\rangle _{T}  &  =\frac{1}{T}\int_{0}%
^{T}F_{N}\left[  q\left(  t\right)  ,p\left(  t\right)  \right]  dt.
\end{align}
Therefore, the non consideration of selfsimilarity could provoke the
persistence of certain states apparently metastable
\cite{lat1,lat2,zanette,dauxois} in a given numerical study, while actually
the characteristic temporal scale of the system evolution has been arbitrarily
dilated, leading in this way to an apparent divergence of the relaxation times.

\section{Conclusions}

Universality of the chaoticity mechanism present in the Hamiltonian dynamics
seems to support the general applicability of a thermo-statistical description
with microcanonical basis in this context, including also the long-range
interacting systems. Therefore, the microcanonical ensemble can be considered
as a safety starting point in order to carry out an appropriate generalization
of the Statistical Mechanics for the Hamiltonian non extensive systems.
Nevertheless, in spite of the ergodic properties of the Hamiltonian dynamics
support a generic characterization in this context, only a deep knowledge
about the features of the microscopic picture leads us to a consequent
characterization of the macroscopic properties of a given system, i.e. the
incidence of the evaporation in the astrophysical systems.

The concept of scaling selfsimilarity have been proposed as an appropriate
generalization of the extensive properties of the traditional systems.
Selfsimilarity is just a general symmetry of the macroscopic description
exhibited under the scaling transformations of the system size. Apparently,
selfsimilarity manifests itself in every macroscopic behavior of the system,
even in the macroscopic dynamics, i.e. the finite character of the relaxation
times under the imposition of the thermodynamic limit compatible with the
system selfsimilarity. Our analysis of Tsallis Statistics \cite{velTsal}, as
well as the study of the thermostatistical characterization of the
astrophysical systems \cite{velastro}, show the relevance of the scaling
selfsimilarity as a sufficient condition in order to justify an appropriate
thermodynamic formalism in the context of the nonextensive Hamiltonian systems.

Most of systems with a practical interest exhibit exponential selfsimilarity
scaling laws, and therefore, the usual Boltzmann-Gibbs statistics should be
applicable in many situations in spite of such systems are dominated by
long-range interactions. An open question in this subject is how to develop a
practical methodology for deriving the specific selfsimilarity scaling
transformation $T_{\alpha}$ acting on those macroscopic observables
determining the macroscopic state of the system. Such a question will be
considered in our forthcoming paper.

\end{document}